# Giant Radio Galaxies – old long-living quasars?

B.V. Komberg[1], I.N. Pashchenko[2]

*Astro Space Center, Lebedev Institute of Physics of Russian Academy of Science, 117997 Moscow, Russia*

## Abstract

Based on the miscellaneous published radio and optical data, SDSS and APM catalogue we consider the various properties of the giant radio sources (gRS) with the aim of refining the conditions leading to the formation of these objects. We compare gRSs with the regular-sized radio sources in radio and optical bands, yielding the following results:

1. The fraction of broad line objects among gRSs with high excitation spectrum is the same as for the RSs from isotropic samples. According to "Unified Scheme" this leads to the isotropic angle distribution of gRSs jets, thus gRSs cannot be characterized as objects with jets lying in the plane of sky.

2. gRSs do not differ from normal sized RSs in apparent asymmetry distribution of their extended radio components (ERC). However the fact that asymmetry distributions for gRSs and giant radio quasars (gQSS) are essentially the same leads within the Unification Scheme to the conclusion that the origin of this asymmetry is in the non-uniform environment.

3. The observed radio jet powers for gRSs and regular RSs are almost the same, so this can hardly explain why gRSs are that large.

4. The richness of the environment for gRSs is the same as for normal sized RSs. Host galaxies of gRSs can be either isolated or be a part of a clusters up to Abell Richness Class 1. This contradicts the opinion that the low density of the environment is the single reason for gRSs formation.

5. Relatively large fraction of Double-Double radio sources among gRSs presumes their lifetimes larger than normal sized RSs by order of magnitude. This fact and coincidence of gRSs space density with space density of Fanaroff-Riley II RSs in local ($z < 0.1$) Universe suggest that about 10% of FRII RSs have by order of magnitude longer lifetimes and eventually evolve to gRSs.

6. The lack of Double-Double gQSSs can be explained by their shorter active phase in comparison with gRGs. In the alternative (to the unification scheme) evolution scheme, which combines radio-loud QSSs and RGs together, former evolve in time into latter. According to this scheme the observed relative quantity of radio quasars in gRSs population (~0.1) can be interpreted as the presence long-living population of radio loud QSSs as ~10% of all radio quasars. Such population of long-living radio quasars can appear to be the parent population for gRGs

## 1. Introduction

Giant radio sources with extended radio component (ERC) sizes $D > 1\,Mpc$ ($h = 0.5$) were discovered in 1974 [1]. Two giant radio sources (3C236 with $z \sim 0.1$ and DA 240 with $z \sim 0.04$) were found to be giant radio galaxies (gRG) with sizes 5 and 2 Mpc respectively. The study of this rare class of radio sources (nearly 140 gRS are known to date [2-8] with redshifts $z \leq 1.8$, $\bar{z} = 0.3$ and angular sizes up to nearly 1 degree) appears to be very interesting due to several reasons. First, studies of gRSs could reveal how RS evolves in time and what physical parameters govern radio source evolution. Second, the possibility of using ERCs as "probes" of intergalactic medium (IGM) arises thanks to their giant linear sizes. The question of possible connection between the prolonged ERC and the distribution of nearest galaxies is also interesting. Furthermore, large angular sizes of gRSs make their significant contribution to small scale (~ ') anisotropy of cosmic microwave background possible via Zel'dovich-Syunjaev effect on relativistic electrons in ERCs of gRSs [9,10].

---

[1] bkomberg@asc.rssi.ru
[2] in4pashchenko@gmail.com

However it is still unclear what exactly reason/reasons lead to gRS formation. It could be special external conditions (low density of IGM [11]) or exceptional internal properties of gRS central engine (high jet power or long activity time [12]). It is likely that none of the mentioned reasons is sufficient itself and several conditions must be actually satisfied [13].

Throughout the paper, the values adopted for cosmological parameters are $\Omega_M = 0.27$, $\Omega_\Lambda = 0.73$ and $H_0 = 71 \; km/(s \cdot Mpc)$ unless stated otherwise.

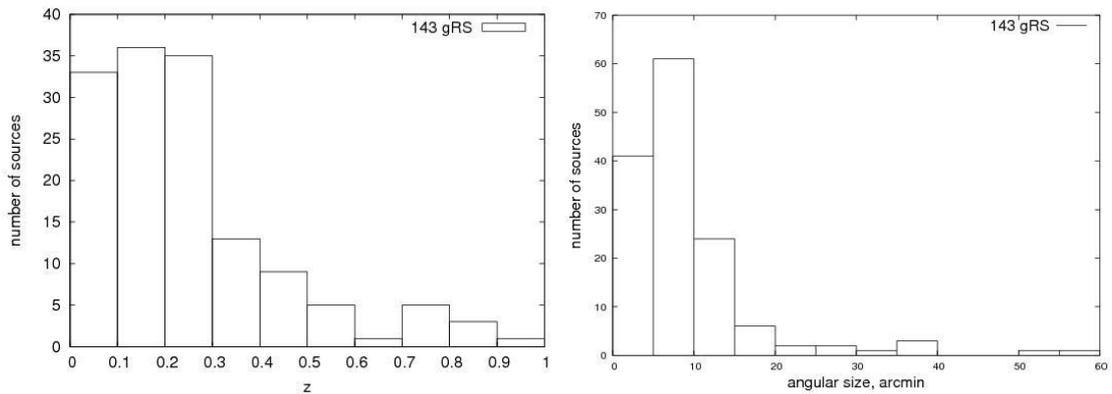

Figure 1. Redshift and angular size distribution of known gRSs from [5],[6]. 2 gRSs in the left figure lie at $z > 1$ (J1432+1548 with $z = 1.005$ and J0903+3943(4C39.24) with $z = 1.883$)

2. Basic properties

1. Here we list some basic properties of gRSs (see [5]). Majority of gRSs are Fanaroff-Riley II [14] RSs as morphologically, as by their radio power $(P_{1.4GHz} \sim 10^{24 \div 28} \; W/Hz)$ - see Fig.1, although nearly ten FRI gRGs and as much intermediate FRI/II gRG are known. Spectral studies of nearly half of gRSs in optical band [15-19] reveal that about half of them are low excitation radio galaxies (LERG) [15, 20, 21] or objects with host galaxies displaying typical spectrum of elliptical galaxy with absorption lines. Nearly third of gRSs are classified as radio galaxies with narrow emission lines/ high excitation radio galaxies (NLRG/HERG) and ~ 20% of gRSs show broad emission lines and classified as quasi stellar sources (QSS) or broad line radio galaxies (BLRG) accordingly with their optical luminosity.

2. It is interesting to examine gRSs properties in terms of Unified Scheme [22, 23]. Authors of [24] obtained fraction of QSSs among all RSs $f_{QSS} = 0.29$ for 3CRR sample of RSs. For gRSs with spectral data from sample of 125 gRSs [5] we get $f_{QSS} = 0.14 \pm 0.04$. However, excluding LERG objects from analysis brings it to $f_{QSS}^{HERS} = 0.27 \pm 0.09$ (that is fraction of QSSs among high excitation RSs - HERSs, considered to include all HERGs and all broad line objects). The fraction of broad line objects among HERS for 3CRR, 6C and 7C samples is $f_{BL}^{HERS} = 0.40$ [21] – and for gRS sample [5] we obtained $f_{BL}^{HERS} = 0.43 \pm 0.11$. *Thus the fraction of broad line objects among giant HERSs is in agreement with isotropic one, based on low frequency flux samples. That is, spectroscopic properties of giant HERSs don't differ from normal sized HERSs.* If the selection effects to large inclination angles of jets in the gRS's sample were significant, there would be deficit of broad line objects. There is opposite result on lower fraction of QSS sources in gRS sample from [6], where only 2 of 18 objects are QSSs. However if one takes QSS fraction among HERSs, then it becomes $f_{QSS}^{HERS} = 0.3 \pm 0.2$.

3. Note that gRSs with FRI morphology are also observed ("classical" FRI RSs with turbulent jets such as NGC 315, 3C 31, 3C 129, 3C 130, J0508+6056, J0918+3151, J1032+5644 HB13, or intermediate "FRII-like" FRI RS with "hot spotless" ERC: J0926+653, J0939+740 sometimes classified as FRII RSs on the basis of their radio power). Jets of "classical" FRI RSs, being relativistic on scales of parsecs are decelerated in external medium on the scales of $\sim kpc$. Thus, growth of such turbulent jet to the length scales $\sim Mpc$ seems to be inconceivable. More realistic assumption include primodal propagation of jet in powerful FRII regime. Subsequent transition to FRI regime could be the consequence of accretion rate reduction. Although physical reasons of Fanaroff-Riley dichotomy remain unclear, presence of giant FRI RSs may suggest that FRI RSs are the remains of far evolved FRII RSs, at least in relatively rich environment (see Section 4). Incidentally, heavier super massive black holes (SMBHs) in FRI RSs [25, 26] also suggest their longer lifetimes.

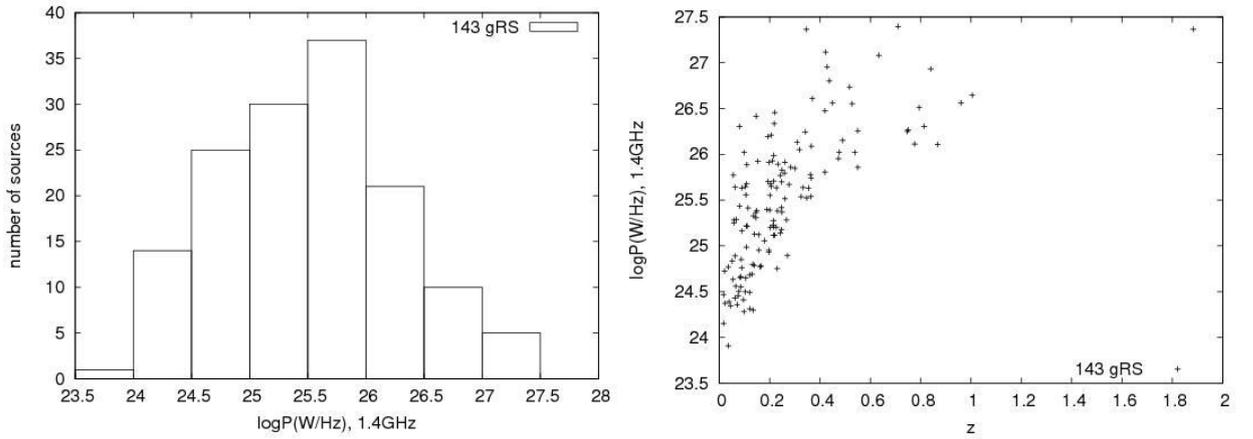

Figure 2. Distribution of total $k$-corrected radio power at 1.4 GHz and $P - z$ diagram for 143 gRSs from [5], [6].

## 3. Asymmetry of extended radio components

The problem of extended radio component asymmetry in radio sources has been the subject of much study (see reviews in [27, 28]). Asymmetry in gRS, it's reasons and connection with asymmetry in normal sized RSs raise the essential interest that is the object of analysis in many papers. In [2] authors claim that gRSs are not more symmetrical then normal sized 3CR RSs (see Figure 12 in cited paper). In [29] the result was obtained that there is no difference in distributions of separation ratios of ERCs for giant and normal sized RSs. However in [30] authors reveal a tendency for gRSs being slightly more asymmetrical then 3CR RSs sharing the same range in redshift. Thus it is not completely clear if gRSs have more symmetrical/asymmetrical ERCs then normal RSs.

We select 70 gRSs with asymmetry parameters known from literature (all of them are FRII RSs) and compare them with asymmetry parameters of 42 3CRR RSs with sizes **$50 < D < 1000\ kpc$** and z<0.6 [31]. Distributions of asymmetry parameter R (which is size ratio of smaller to longer ERC) for both samples are presented in Fig.1. The mean R for gRS sample is $\bar{R}_{gRS}$ **= 0.76 ± 0.02**, and for 3CRR sample is $\bar{R}_{3CRR}$ **= 0.80 ± 0.03**[3]. Kolmogorov-Smirnov test doesn't reveal significant difference in distribution of any asymmetry parameter $R$, $x =$ **$(1 - R)/(1 + R)$** or misalignment angle C (null hypothesis that studied distributions are selected from one distribution of $R$, $x$ and $C$ can be rejected only at ~0.2, 0.3 and 0.5 level of significance).

However, one can discern the reason of asymmetry in gRSs by examining asymmetries of QSSs and RGs in both samples. Distributions of asymmetry parameter $R$ in 3CRR QSSs and RGs differ significantly – Mann-Whitney U-test rejects null hypothesis that both distributions are sampled from one

---

[3] $\bar{R}_{3CRR}$ **= 0.81 ± 0.02** without RG 3C299, high level of asymmetry $R$ **= 0.31** in this RG caused by interaction of one jet with high density gas cloud, clearly seen in emission lines [32]

at significant level 0.5 %, and after adding 4 RGs and 1 QSS[4] with **50 < $D$ < 1000** $kpc$ and **0.6 < $z$ < 0.75** - at 0.2%. This result on whole 3CRR sample was considered in [31] as QSS's jets being close to the line of sight, that agrees with Unification Scheme. In light travel time model ERC asymmetry manifests itself exactly just at smaller inclination angles so authors of [31] showed that the main reason of ERC asymmetry in normal sized 3CRR RSs is geometrical effect (light travel time delay). However distributions of asymmetry parameter $R$ for two optical classes of gRSs (58 gRGs and 11 QSSs) statistically do not differ. It suggests that geometrical delay is not the main reason of ERC asymmetry formation in gRSs. Thus one infers that non-uniform external medium is responsible for asymmetry formation in gRSs.

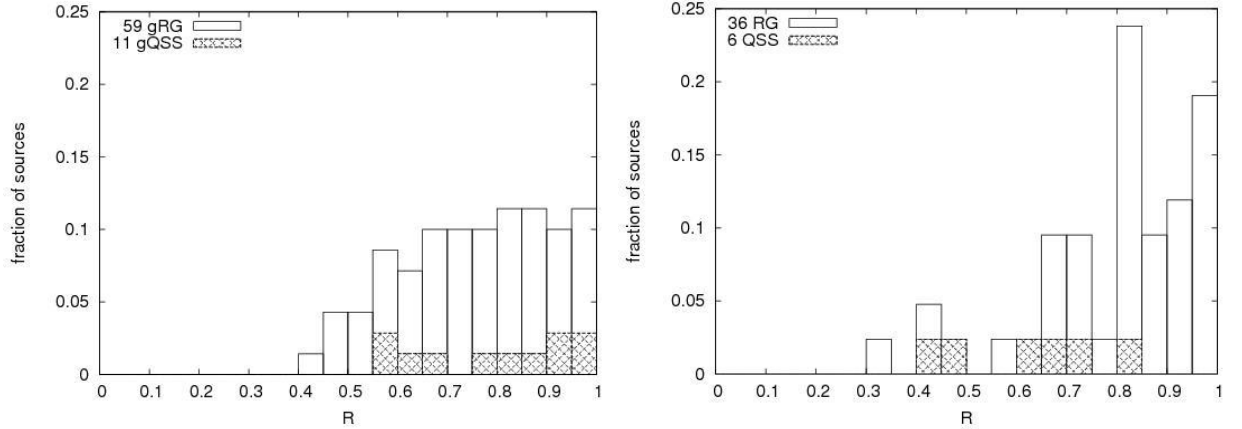

Figure 3. Distributions of asymmetry parameter $R$ for 70 gRSs (left) and 42 3CRR RSs with **50** $kpc$ < $D$ < **1000** $kpc$ and $z$ < **0.6** (right). Hatched bins represent the QSSs distribution.

### 4. Jet Power and host galaxies of gRSs.

1. GRSs could reach the large sizes as a result of high jet powers $P_{jet}$ and hence high advance speeds of jets through ambient medium. Unfortunately, $P_{jet}$ is hard to estimate because of the absence of any full reliable jet model up to date. Recently a method of estimating $P_{jet}$ through the amount of mechanical work that ERCs do while blowing "bubbles" in hot gas of ambient medium of host cluster has appeared [35, 36]. Though this method is free from most of jet model uncertainties, it is inapplicable for gRSs because of their avoidance of rich clusters (see 5). If one assumes that $P_{radio}$ (that is energy of relativistic electrons/positrons and magnetic fields) is proportional to $P_{jet}$ and conversion efficiency of $P_{jet}$ in energy of radiating particles is constant for gRSs and normal sized RSs then it is possible to compare $P_{radio}$ for both classes of RSs and test the scenario of gRSs attaining their large sizes through high jet powers. Authors of [37] studied VLBI observations of gRSs DA 240 and J1331-099 and showed that Doppler-corrected radio power is $P_{radio}$~$10^{24 \div 25} W/Hz$, which doesn't exceed $P_{radio}$ of regular sized RGs and QSSs or is even less if slower moving "sheath" is observed. Substantial disadvantage of this analysis is that VLBI observations probe the current gRSs nuclear activity, but $P_{jet}$ could be higher in the early stage of activity.

2. One of the possible reasons for hypothetically high $P_{jet}$ in gRs could be the high SMBH mass in the center of their host galaxies. Connection is expected between $P_{jet}$ and SMBH mass in all of the models of jet formation (see for example [38, 39]). Furthermore, the authors of [40] (see also [41]) reveal dependence of radio loud active galactic nuclei (AGN) fraction on the host galaxy mass as $f_{RL}$~$M_*^{1.8}$. Being interpreted in the key that AGNs in the massive host galaxies stay active for the longer

---
[4] without QSS 3C254, high asymmetry of which caused by interaction of jet with high density gas clouds within host galaxy [33] and RG 3C441, high asymmetry of which is developed through interaction with neighbor galaxy [34]

times or more frequently come to active phase of recurrent activity [41], it suggests that gRSs formation is caused by long lifetimes of RS activity in massive galaxies. However, longer lifetimes of RS activity also imply large SMBH masses just as a result of longer accretion times. It is easy to verify this suggestions studying the properties of gRS host galaxies since SMBH mass is connected with mass of elliptical galaxy that hosts RS [42] or it's optical luminosity [43] etc. This analysis will be done in the subsequent paper.

## 5. Environment of gRSs.

There is a lack of any direct study of gRSs environment. Nevertheless, authors of [5] claim that majority of gRSs avoids the rich clusters. Thus there is sense in studying the environment of gRSs in more detail. The point is that some authors consider low density of ambient medium [7] or even rich optical environment without dense X-Ray emitting medium [44] (that implies relatively short dynamical age of group/cluster) as a possible reason of gRS formation.

We conduct the study of optical environment of $z < \sim 0.1$ gRSs. We used SDSS [45] data where it was possible to count the nearest galaxies. For gRSs lying off SDSS area we used APM catalog [46] for obtaining amplitudes of two point correlation function (CF) $B_{gg}$ [47]. CF amplitudes corresponding Abell Richness Class $N_{ARC} = 0/1$ are $322 \pm 108/537 \pm 108$ $Mpc^{1.77}$ [48]. The mean $B_{gg}$ for the galaxies from several galaxy catalogs is $B_{gg} \approx 29$ $Mpc^{1.77}$ [49]. Among the sample of 125 gRSs from [5], 56 objects fall into the area of SDSS Data Release 7. We choose 16 objects with the magnitude $m_r$ corresponding to the galaxy of characteristic absolute magnitude $M_r^*$ lying at the redshift of target object, brighter then $m_r = 17.77$, where $M_r^* = -21.37$ and all parameters of galxy luminosity function are taken from [50]. Redshifts and coordinates of host galaxies are taken from [2,3,4,5], all transformations between optical bands for type "E+S0" galaxies are taken from [51]. Table 1 presents results for SDSS sample. These are: object's name in standard (J2000) and SDSS format, Fanaroff-Riley type, redshift, $m_r$- apparent magnitude in SDSS r band, «$N_{0.5}$» - number of galaxies within projected distance 0.35 Mpc (0.5 Mpc for h=0.5) and velocity $\Delta v = \pm 600$ $km/s$ of gRS host galaxy, «$N_{0.5}^{-19}$» - the same as «$N_{0.5}$», but galaxies brighter then $M_V = -19$ and its error calculated as «$N_{0.5}^{-19}$» $/\sqrt{\text{«}N_{0.5}\text{»}}$, $B_{gg}$ – amplitude of CF and its error calculated as $\sigma_{B_{gg}} = B_{gg}/\sqrt{N_{0.9}}$.

Table 1. SDSS based environment measures for 16 gRSs

| J2000 | SDSS | FR | z | $m_r$ | «$N_{0.5}$» | $N_{0.9}$ | «$N_{0.5}^{-19}$» | $B_{gg}$, $Mpc^{1.77}$ |
|---|---|---|---|---|---|---|---|---|
| 0508+6056 | J050827.24+605627.5 | I | 0.071 | 15.97 | 0 | 0 | 0 | - |
| 0636-2034 | J063632.25-203453.3 | II | 0.056 | 15.11 | 0 | 0 | 0 | - |
| 0918+3151 | J091859.40+315140.6 | I | 0.062 | 14.63 | 7 | 12 | 8 ± 3 | 209 ± 60 |
| 1006+3454 | J100601.73+345410.5 | II | 0.0992 | 15.36 | 0 | 2 | 0 | 75 ± 53 |
| 1032+2756 | J103214.01+275601.6 | II | 0.085 | 15.76 | 1 | 1 | 2 ± 2 | 28 ± 28 |
| 1032+5644 | J103258.88+564453.2 | I | 0.045 | 13.55 | 10 | 33 | 8 ± 3 | 399 ± 69 |
| 1113+4017 | J111305.54+401729.8 | I/II | 0.0745 | 14.90 | 4 | 29 | 6 ± 3 | 649 ± 121 |
| 1147+3501 | J114722.13+350107.5 | II | 0.063 | 14.53 | 8 | 10 | 9 ± 3 | 179 ± 56 |
| 1247+6723 | J124733.31+672316.4 | II | 0.1073 | 16.03 | 0 | 0 | 0 | - |
| 1311+4059 | J131143.08+405859.7 | II | 0.1105 | 16.07 | 1 | 3 | 3 ± 3 | 149 ± 86 |
| 1312+4450 | J131217.00+445021.2 | I | 0.0358 | 13.35 | 8 | 16 | 5 ± 2 | 161 ± 40 |
| 1328-0307 | J132834.36-030744.7 | II | 0.0852 | 16.70 | 0 | 1 | 0 | 29 ± 29 |
| 1418+3746 | J141837.65+374624.5 | II | 0.1349 | 16.32 | 1 | 6 | 6 ± 6 | 516 ± 211 |
| 1428+2918 | J142819.23+291844.2 | II | 0.087 | 15.37 | 1 | 5 | 2 ± 2 | 145 ± 65 |
| 1552+2005 | J155209.19+200523.2 | II | 0.090 | 15.90 | 1 | 9 | 2 ± 2 | 304 ± 101 |
| 1628+5146 | J162804.05+514631.3 | II | 0.0547 | 14.90 | 0 | 0 | 0 | - |

After that we chose all gRSs with $z < 0.1$ and $|b| > 30°$ that missed SDSS (20 objects) and evaluated amplitudes of CF using APM catalog. Details on this procedure can be found in [52], where the counting radius and magnitude maximizing signal-to-noise ratio are selected or in [48], where errors arising from different counting radius and magnitude and different normalizing galaxy luminosity function are estimated. We chose counting magnitude which being extinction corrected corresponds to $M_r^* + 1$ on the redshit of gRSs, counting radius 0.9 Mpc and all galaxy luminosity function parameters from [47]. Results are listed in Table 2, where the columns are: J2000 name, alternative name, Fanaroff-Riley type, redshift, $N_{0.9}$ - number of extended objects within projected distance of 0.9 Mpc and with APM extinction corrected E magnitude corresponding to the redshift of gRS being brighter then $M_r^* + 1$, $N_{0.9}^{corr}$ - the same, but background corrected by subtracting counts from several nearest areas, $B_{gg}$ – amplitude of CF and it's error calculated as $\sigma_{B_{gg}} = B_{gg}/\sqrt{N_{0.9}^{corr}}$.

Table 2. APM based environment measures for 20 gRSs.

| J2000 | Alt. name | FR | z | $N_{0.9}$ | $N_{0.9}^{corr}$ | $B_{gg}, Mpc^{1.77}$ |
|---|---|---|---|---|---|---|
| 0057+3021 | NGC 315 | I/II | 0.0167 | 30 | 7.8 | 230 ± 81 |
| 0107+3224 | 3C 31 | I | 0.0169 | 60 | 35.0 | 1148 ± 184 |
| 0320-4515 | | II | 0.0633 | 11 | 3.5 | 145 ± 78 |
| 0505-2835 | | II | 0.038 | 3 | 1.5 | 69 ± 56 |
| 0513-3028 | | II | 0.0583 | 2 | -1 | −56 ± 56 |
| 0749+5554 | DA 240 | II | 0.0356 | 11 | 5.3 | 397 ± 173 |
| 0918+3151 | | I | 0.062 | 12 | 6 | 244 ± 99 |
| 0949+7314 | 4C 73.08 | II | 0.0581 | 1 | -5 | −300 ± 134 |
| 1018-1240 | | I/II | 0.0777 | 9 | 0.5 | 23 ± 32 |
| 1032+2756 | | II | 0.0854 | 3 | -3.8 | −185 ± 95 |
| 1032+5644 | HB 13 | I | 0.045 | 15 | 9.5 | 552 ± 179 |
| 1113+4017 | | I/II | 0.0745 | 33 | 26.8 | 855 ± 165 |
| 1147+3501 | | II | 0.0630 | 8 | 2.8 | 249 ± 150 |
| 1312+4450 | | I | 0.0358 | 17 | 7.8 | 391 ± 140 |
| 1328-0307 | | II | 0.0860 | 2 | -3.3 | −267 ± 148 |
| 1334-1009 | | II | 0.081 | 7 | 0.3 | 19 ± 37 |
| 1428+2918 | | II | 0.087 | 17 | 5 | 215 ± 96 |
| 1552+2005 | 3C 236 | II | 0.0895 | 33 | 24.5 | 493 ± 100 |
| 1628+5146 | | II | 0.0547 | 3 | -3.3 | −238 ± 129 |
| 1632+8232 | NGC 6251 | I/II | 0.023 | 1 | -2.3 | −647 ± 431 |

One can conclude that gRSs populate quite different environment: host galaxies of some gRSs are virtually isolated systems although some of them can be the part of clusters up to clusters of Abell Richness Class $N_{ARC} = 0$ (gRS of intermediate FR type I/II - "fat double" J1418+3746 with $B_{gg} = 516 ± 211\ Mpc^{1.77}$ residing in Rood-Sastry [54] class "L" [55] cluster Abell 1896; FR type I gRSJ1032+5644 HB13 with $B_{gg} = 399 ± 69\ Mpc^{1.77}$ residing in cluster №3047 of SDSS C4 Galaxy Cluster Catalog (DR3) [56]) or even $N_{ARC} = 1$ (FR type II gRS J1113+4017 with $B_{gg} = 649 ± 121\ Mpc^{1.77}$ residing on the outskirt of Abell 1203 cluster with z=0.00751 and velocity dispersion $\sigma_v = 296\ km/s$ [53], [57]). Interestingly, in the paper [58] a lower limit on X-Ray luminosity from Abell 1203 was obtained: $L_X < 0.37 \cdot 10^{43}\ erg/s$, that matches well with it's classification of Rood-Sastry type "C" and Bautz-Morgan class [59] "II-III" [55] that implies its relatively short dynamical age. That agrees well with low ($L_X \sim 10^{42} h^{-2}\ erg/s$) X-Ray luminosity of nearby (z<0.35) clusters with $N_{ARC} = 0 ÷ 3$ hosting FR type II RS that was found in [60]. However, the majority of studied gRSs (being mainly

FR type II RSs) inhabit environment that agrees with groups of a few members coinciding with environment of near FR type II RSs studied in [52], [61].

*Summing up, one can claim that environment of gRSs fits well with general picture according to which the majority of nearby FR type II RSs inhabit poor groups but can be the part of clusters up to clusters of $N_{ARC} = 1$ and higher, however, lacking significant X-Ray emission. Thus, FR type II RSs reside in dynamically young systems [60], [61], and FRI RSs inhabit dynamically evolved richer environment [61]. This can imply an absence of any single condition for gRS formation. In other words formation of different individual gRSs can be caused by different reasons: as low ambient medium density, as long activity time of RS, as relatively high jet power [13].*

## 6. Long lifetime as a reason of gRS formation

1. Essential questions arise: could gRSs be far evolved FR type II RSs and can any FRII RS form the gRS? These questions are especially relevant because as it can be seen from discussion above the gRSs don't seem to be different from the normal sized RSs neither in richness of their environment (see Section 5), nor in jet power (see Section 4.1). In paper [6] the space density of nearest ($z < 0.13$) gRSs was obtained - $n_{gRS} = (1.1 \pm 0.6) \cdot 10^{-7} Mpc^{-3}$. The space density of nearest FR type II RSs is $n_{FRII\ RS} \approx 2.3 \cdot 10^{-7} Mpc^{-3}$ [62]. In $z < 0.1$ region of WENSS gRS sample where selection effects play no significant role the 18 gRSs are located so one obtains $n_{gRS} = (3.1 \pm 0.7) \cdot 10^{-7} Mpc^{-3}$ (high value of space density from WENSS data may result from the higher WENSS sensitivity and large sample size).

Thus if RS lifetime in giant phase doesn't differ from lifetime in normal size phase of activity then the coincidence of the space densities implies that each FR type II RS evolves in gRS (at least in near Universe).

However one can't reject the possibility that gRS formation is due to the long activity time of the "central engine" or due to recurrent activity at some part of normal sized RS population being the parent population for gRSs. Indeed, authors of [12] suggest the recurrent activity in gRSs basing on some peculiarities of ERCs in few gRSs, namely hotspots well recessed from the ends of lobes and jet discontinuities. Radio sources with morphological signs of recurrent activity are also 4 of 12 gRSs from SUMSS sample [6]. We suppose that more significant evidence in favor of long activity timescale in gRSs is the relatively large number of radio galaxies with two pairs of ERCs (Double-Double Radio Galaxies) [65]: older outer and younger inner pairs. That is 9 gRSs among 13 known DDRGs [8], [64].

If DDRG phenomenon is considered to reflect discontinuity in beam production then it is possible to infer the $\tau_{gRS}/\tau_{RS}$ ratio of gRS and normal RS lifetimes from relative number of Double-Double radio sources in gRS and normal sized RS populations. If one assumes that typical time $\tau_{DD}$ between events triggering jet interruption (the period of time during which the interruption of jet activity is possible) is the same for both populations (that is supported by conclusions of Section 5) then the relative number of Double-Double objects in populations is $\tau_{gRS}/\tau_{DD}$ and $\tau_{RS}/\tau_{DD}$ for gRSs and normal sized RSs respectively. As follows from [65], observations give the number of gRSs ~ 10% of FR type II normal sized RSs. Thus fraction of Double-Double objects in gRS and normal FRII RS population is $f_{gRS}^{DD} = 9/143 \approx 6\%$ and $f_{RS}^{DD} = 4/10 \cdot 143 \approx 0.3\%$. Then $\tau_{gRS}/\tau_{RS} = 20$, that agrees well with results of FRII RS evolution modeling [9]. This is also coincides with ratio of synchrotron age of giant ERCs: $\tau_{syn} \sim 10^8$ yr to kinetic age of normal sized RS: $\tau_{kin}[yr] = 3 \cdot 10^7 D[Mpc]/\beta[0.1c]$. Moreover, it is also agrees with the fact of Triple-Triple radio galaxy discovery in gRS population [66]: $(\tau_{gRS}/\tau_{DD})^2 \approx 1/143 \sim 10^{-2}$ or $\tau_{gRS} \sim 0.1 \tau_{DD}$, that coincides with estimation from $f_{gRS}^{DD} = \tau_{gRS}/\tau_{DD} \sim 0.1$. Taking $\tau_{gRS} \sim 10^8$ years, one gets the characteristic time $\tau_{DD} \sim 10^9$ лет. This coincides with the time scale of galaxy merging due to dynamical friction [67], which supports the hypothesis of the formation of the DDRS formation via jet interruption.

*Thus, combining the facts of equal gRS and FR type II normal sized RS space densities in nearby Universe with derived ratio of gRS to normal RS lifetimes one concludes that ~1/10 part of FRII RS population has for some reason longer lifetimes (by the order of magnitude) and consequently could evolve in gRSs.*

2. The problem of activity time length in radio band is of utmost importance due to the existence of gQSSs that constitute ~10% to the known gRS population. Therefore one can suppose that ~ 10% of radio quasars have an order of magnitude longer phase of activity and hence become gQSSs which, in turn, can evolve to gRSs[5]. Such point of view on quasar phenomena contradicts to standard Unified Scheme. Indeed, the expected (see section 2) number of gQSS objects among DDRGs is **(0.9 ± 0.6) ÷ (1.7 ± 0.9)**[6]. However, there is no quasars among 9 known DDRGs (this justifies using the term "DDRG"). That fact can be due to shorter lifetimes of gQSSs relative to gRGs and it agrees with observed statistics: there are 12 gQSSs observed among spectroscopically classified objects of gRS sample. So ~ 24 objects are expected among full sample (under assumption of uniform distribution of Double-Double objects among optical classes of gRSs there would be 1.5 gQSSs and 7.5 gRGs with Double-Double morphology). As there is no QSSs among DD objects the fraction is $\boldsymbol{f_{gQSS}^{DD}}$ **< 1/24 ≈ 4%** and for RGs $\boldsymbol{f_{gRG}^{DD}}$ **= 9/(143 − 24) ≈ 8%**. Then $\tau_{gQSS}/\tau_{gRG}$ **< 0.5** which agrees with our proposition of evolution from long-lived gQSSs to gRGs. Of course all this conclusions must be checked with larger samples of gRSs.

This research has made use of the NASA/IPAC Extragalactic Database (NED) which is operated by the Jet Propulsion Laboratory, California Institute of Technology, under contract with the National Aeronautics and Space Administration.

Funding for the SDSS and SDSS-II has been provided by the Alfred P. Sloan Foundation, the Participating Institutions, the National Science Foundation, the U.S. Department of Energy, the National Aeronautics and Space Administration, the Japanese Monbukagakusho, the Max Planck Society, and the Higher Education Funding Council for England. The SDSS Web Site is http://www.sdss.org.

The SDSS is managed by the Astrophysical Research Consortium for the Participating Institutions. The Participating Institutions are the American Museum of Natural History, Astrophysical Institute Potsdam, University of Basel, University of Cambridge, Case Western Reserve University, University of Chicago, Drexel University, Fermilab, the Institute for Advanced Study, the Japan Participation Group, Johns Hopkins University, the Joint Institute for Nuclear Astrophysics, the Kavli Institute for Particle Astrophysics and Cosmology, the Korean Scientist Group, the Chinese Academy of Sciences (LAMOST), Los Alamos National Laboratory, the Max-Planck-Institute for Astronomy (MPIA), the Max-Planck-Institute for Astrophysics (MPA), New Mexico State University, Ohio State University, University of Pittsburgh, University of Portsmouth, Princeton University, the United States Naval Observatory, and the University of Washington.

We acknowledge the use of APM catalog. The APM Web Site is http://www.ast.cam.ac.uk/~apmcat/

---

[5] of course the properties of such long-living quasars must differ somehow from normal radio loud quasars. This is still an open issue. Apparently, the reason could be either kinematic properties of host group, either peculiarity in accretion process as it proposed in [68], or more slower decline of SMBH angular momentum with time [69], [70], [71]. It may also be that there is no single reason

[6] the exact value depends on unknown spectral classification of J0041+3224 and J0116-4722. The cited values correspond to both LERG and HERG objects respectively